\documentclass[graybox]{svmult}

\usepackage{amsmath,amssymb,amsfonts}
\usepackage{mathptmx}       
\usepackage{helvet}         
\usepackage{courier}        
\usepackage{type1cm}        
\usepackage{makeidx}         
\usepackage{graphicx}        
\usepackage[colorlinks]{hyperref}
\usepackage{multicol}        
\usepackage[bottom]{footmisc}

\def\sC{\mathcal{C}}
\def\sL{\mathcal{L}}

\def\ssC{\scriptstyle C}

\def\dsC{\mathcal{C}^\dagger}

\newcommand{\state}[1]{|#1\rangle}

\newcommand{\be}{\begin{equation}}
\newcommand{\ee}{\end{equation}}
\newcommand{\beq}{\begin{equation}}
\newcommand{\eeq}{\end{equation}}
\newcommand{\bea}{\begin{eqnarray}}
\newcommand{\eea}{\end{eqnarray}}

\newcommand{\eps}{\epsilon}

\makeindex             

\begin{document}

\title*{Stability longevity and all that : false vacua and topological defects}
\author{Urjit A Yajnik}
\institute{Urjit A Yajnik \at Indian Institute of Technology Bombay, Powai Mumbai 400076, \email{yajnik@iitb.ac.in}
}
%
%
\maketitle

\abstract{I present two interesting studies related to the role of solitons in 
theories with spontaneous symmetry breaking. Quantised 
fermions coupled to solitons are known to induce fractional fermion number. I 
present an example where an unstable topological solution binding a zero 
mode of a majorana fermion gets stabilised due to the induced fractional 
number. Thus neither is the cosmic string stable nor is the fermion number 
conserved, yet the bound state then becomes stable due to Quantum Mechanics. 
The other phenomenon concerns a metastable vacuum getting destabilised due to 
the 
presence of a cosmic string. This happens because the scalar field signalling 
spontaneous symmetry breaking acquires a vacuum expectation value approaching 
its true vacuum value in the core of the string. Thus the string acts like a 
seed nucleating the true vacuum. The instanton mediating between the false 
vacuum and the exit point into the true vacuum is shown to exist and graphically 
displayed through a numeric calculation. T. Padmanabhan has been a long time 
friend and was collaborator for early investigations in this direction. I review 
these works done subsequently with other collaborators as a felicitation to him 
on his sixtieth birthday.}

\section{Cuisine and a canine}
\label{sec:cc}
I got to know Paddy during the last year of MSc at IIT Bombay as my senior Kandaswamy had joined 
PhD in Prof. Jayant Narlikar's group, the same group as Paddy. Paddy already had a reputation of sorts as a very sharp student and a friendly guy. 
I ended up going to Texas Austin for PhD, and as luck would have it, he showed up there in the last year of my PhD. He had been invited
for a visit by my advisor Prof. E. C. George Sudarshan, who was then busy 
reinvigorating Institute of Mathematical Sciences, Chennai known as Matscience 
in those days, 
as its new Director. 

Paddy took interest in the work I had just  then wound up for my thesis. One of the papers dealt with phase transition induced by cosmic strings
\cite{Yajnik:1986tg}.
I had identified the mechanism qualitatively and set up an example to demonstrate it, and was able to simulate the false vacuum decay.
These were very interesting possibilities to explore as the dynamics of inflationary Universe was then believed to rely on the availability
of a metastable or false vacuum. However there was a need to make the detailed mechanism explicit. Elegant arguments such as demonstrating 
the existence of an instanton were not working, that is, proving too difficult. Paddy who was not himself working in any of those topics listened carefully 
and made a pragmatic  suggestion for my open problem. That was to was to analyse the  small oscillations and see the emergence  of a zero 
frequency mode as the temperature was lowered. This worked out nicely and appeared as ref. \cite{Yajnik:1986wq}. 

Out of general friendship, and because he was a visitor to my Centre, and because he was also a senior, but perhaps also because he was making 
such fruitful suggestions for  research, it was only appropriate that I invited him home once in a while.  His apartment was nearby but then some 
neighbours used to leave their dog free to roam at night, causing understandable concerns for Paddy especially on a short visit in a foreign 
country. So I also used to drop him off  by car. It was during these sessions that he ended up discovering that I was reasonably good at making 
pasta sauce and for that reason, I assume among others, he came to decide that I could be a good addition as a visiting fellow to his group in TIFR. 
And so the following January, 1987, I joined Theoretical Astrophysics Group as 
a Visiting Fellow. 

I knew him much better over the next few years in TIFR where he was a patient and amused listener  (he still is I believe) of my daily observations 
about the way the world worked  and also continued providing useful comments on the work I was doing. He was during that time as proilific 
as he has ever been. I remember particularly hearing from him about the Dark Matter problem which had not till then pricked the conscience of 
Particle Physicists sufficiently,  and a ``white paper" on inflationary density 
fluctuations, while some ideas that occupied him then included emergence of a 
fundamental length scale at the Planck scale.  It was a very pleasant and 
fruitful time and a time  to make some bold calculations on particle production 
during inflation \cite{Yajnik:1990un}. 
Subsequently some of the undergraduate students from IIT Bombay went and worked with him at IUCAA during summers and we were in touch in that way. 

Paddy has been a wellspring of innumerable new ideas and insights, has moved into all the emerging areas of General Relativity and Cosmology that
seemed to bear on the fundamental issues, written many instructional and pedagogical articles and books, and has marked out a space for India 
in the world in those topics. 
Here I choose to contribute on the problem that was our first and only joint 
work, presenting some interesting further evolution of 
the ideas originated then, with the help of other colleagues as appears in references \cite{Sahu:2004zp} and \cite{Lee:2013zca}.

\section{Introduction}
\label{sec:intro}
In this contribution I cover two works, one in which an otherwise unstable 
topological configuration becomes stabilised and the other, in which a 
topological object can hasten the decay of a false vacuum. While an infinite 
cosmic string can be topologically stable, the same when made into a closed 
loop 
becomes unstable to decay by shrinking. However such strings can bind zero modes 
of fermions, which in some cases requires assigning fractional fermion number to 
the string. Such fermions can be Dirac or Majorana. Majorana fermions do not 
have a conserved fermion number. Yet it is interesting to prove that when a 
Majorana fermion is bound to a cosmic string loop, both configurations by 
themselves unstable, the combination becomes stable to spontaneous decay. This 
is due to the unusual assignment of fermion number of the configuration, whose 
quantum numbers cannot be matched to those of the vacuum. This is shown in 
Sections \ref{sec:topsolfracferno}  
and \ref{sec:ferno}. An 
interesting aspect of this analysis is that instead of an unfathomable 
\textit{Dirac sea} we have a small \textit{Majorana pond}  at the energy 
threshold. The occurrence of a finite number of gapless states degenerate with 
the vacuum has been argued in \cite{mcwilczek} to be a signature of 
spontaneous symmetry 
breaking with a fermionic order parameter.  

In a second study which takes us back to the problem on which I collaborated  
with Paddy, cosmic strings can induce the decay of a false vacuum. In our 
attempts then we showed that  there were reasonable theoretical grounds for the 
phenomenon and proved it by numerical simulations. But in more recent work, 
thanks to the collaborators cited, we were able to doctor the model sufficiently 
that there was more theoretical control. In this case, it is possible to extend 
the well known theoretical tool of an ``instanton bounce" to this case. Unlike 
in a translation invariant false vacuum, quite a bit of more ingenuity is 
required for deducing a bounce in the presence of a topological object, which 
breaks the translation invariance. However when everything is put together one 
gets an explicit dependence of the decay rate of the false vacuum on the 
parameters of the model. We can then see the enhancement in the rate, as also a 
possible regime of instability implied only due to the presence of the 
topological object. While the analytical answer could be obtained in the 
2+1  dimensional case of a vortex \cite{Lee:2013ega}, here we take up the 3+1 
dimensional cosmic string which presents interesting theoretical challenges. In 
this case numerical calculations assist substantially in visualising and 
verifying the bounce. This is shown in Sec.s \ref{sec:falsestring} and 
\ref{sec:instantonandbulge}.

\section{Topological solutions and fractionalised fermion number}
\label{sec:topsolfracferno}
Cosmic strings and vortices are examples of solitons, extended objects occuring 
as stable states \cite{3}\cite{4} within Quantum Field
Theory present the curious possibility of fermionic 
zero-energy modes
trapped on such configurations. Their presence requires,
according to well known arguments\cite{JandR}\cite{Jrev}, an
assignment of half-integer fermion number to the
solitonic states.  Dynamical stability of such objects
was pointed out in \cite{devega}, in cosmological context in \cite{stern},
\cite{steandyaj} and later studied also
in \cite{Davis:1997bu}\cite{Davis:1999wq}\cite{DavKibetal}. 
Fractional fermion number phenomenon also occurs in condensed
matter systems and its wide ranging implications call for
a systematic understanding of the phenomenon.

The impossibility of connecting half-integer valued states 
to integer valued states suggests that a superselection 
rule\cite{www} \cite{Sweinberg} is operative. 
In a theory with a conserved charge (global or local), a 
superselection rule operates among sectors of distinct charge 
values because the conservation of charge is associated 
with the inobservability of rescaling operation $\Psi\rightarrow 
e^{iQ}\Psi$. In the case
at hand, half-integer values of fermion number occur, preventing such states 
from 
decaying in isolation to the trivial ground state \cite{devega}
\cite{stern}. 

\begin{figure}[htb]
{\par\centering \resizebox*{0.3\textwidth}{!}
{\rotatebox{0}{\includegraphics{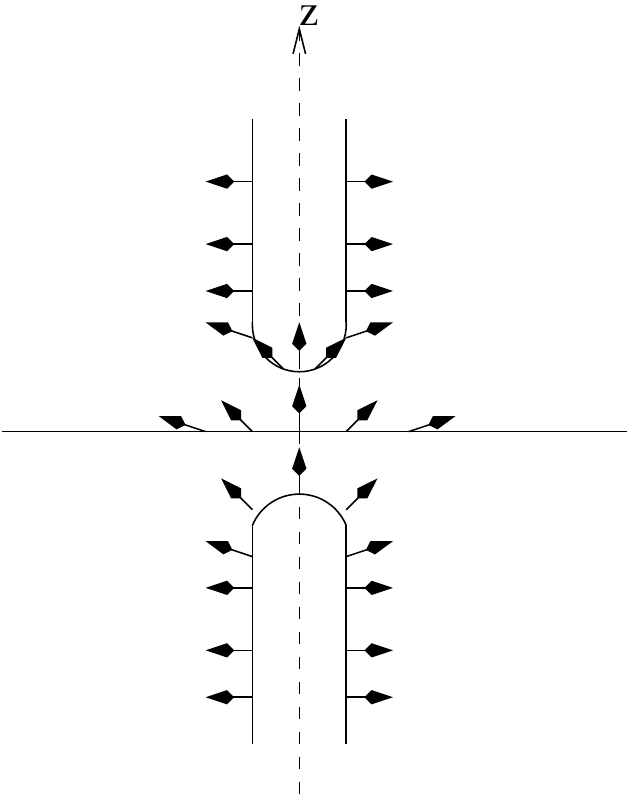}}} \par}\vspace{0.3cm} 
\caption{Schematic configuration depicting break of a string into monopole 
anti-monopole pair. The diagram shows isospin vectors after the
rupture of the string. Internal orientations are mapped to external 
space. They are shown just outside the core of the two resulting 
pieces and on the  mid-plane symmetrically separating the two.}
\label{fig:string}
\end{figure}

Here we construct an example in which the topological object of 
a low energy theory is metastable due to the embedding of the 
low energy  symmetry group in a larger symmetry group at higher 
energy. Examples of this kind were considered in \cite{presvil}.
Borrowing the strategies for bosonic sector from there, we include 
appropriate fermionic content to obtain the required zero-modes.
Consider a  theory  with local $SU(3)$ symmetry broken 
to $U(1)$ by two scalars, $\Phi$ an octet acquiring a VEV $\eta_1\lambda_3$ 
($\lambda_3$ here being the third Gell-Mann matrix) and 
$\phi$, a ${\bar 3}$, acquiring the VEV $\langle\phi^k\rangle=\eta_2
\delta^{k 2}$, with $\eta_2\ll \eta_1$. Thus
\be
SU(3) {\buildrel 8 \over{\longrightarrow}}
U(1)_3\otimes U(1)_8 { \buildrel \overline{3}  \over{\longrightarrow}} U(1)_+
\ee
Here $U(1)_3$ and $U(1)_8$ are  generated by $\lambda_3$ 
and $\lambda_8$ respectively, and $U(1)_+$ is generated by
\( (\sqrt{3}\lambda_8 + \lambda_3)/2 \) and likewise $U(1)_-$
to be used below.  
It can be checked that this pattern of VEVs can be generically
obtained from the quartic scalar potential of the above Higgses. 
The effective theory at the second breaking $U(1)_-\rightarrow \mathbb{Z}$ 
gives rise to cosmic strings. However the $\mathbb{Z}$ lifts to identity in 
the $SU(3)$ so that the string can break with the formation of 
monopole-antimonopole pair. See Fig. \ref{fig:string}

Now add a multiplet of left-handed fermions belonging to $\overline{15}$. 
Its mass terms arise from the following coupling to the $\overline{3}$
\be
\sL_{\textrm{Majorana}} = h_M \overline{\psi^{\ssC}}^{\{ij\}}_{k}
\psi^{\{lm\}}_{n}\phi^{r}
(\epsilon_{ilr}\delta^{n}_{j}\delta^{k}_{m})
\ee
The indices symmetric under exchange have been indicated by curly brackets. 
No mass terms result from the $8$ because it cannot provide a singlet
from tensor product with $\overline{15}\otimes \overline{15}$ 
\cite{Slansky:yr}. 
On substituting the vacuum expectation value (vev) of $\phi$ we get the mass 
matrix $M$ of the fermions. A systematic 
enumeration shows that all but the two components $\psi^{\{22\}}_{1}$ 
and $\psi^{\{22\}}_{3}$ acquire
majorana masses at the second stage of the breaking. Specifically 
we find the majorana mass matrix to be indeed rank $13$. In the cosmic 
string sector, the vev of $\phi$ and therefore the mass matrix $M$ becomes 
space dependent, $~e^{i n \theta}$, where $\theta$ is angle in a plane
perpendicular to the string, and $n$ is the winding number which has to be 
integer.
The lowest energy bound states resulting from this coupling
are characterized by a topological index, \cite{Eweinberg}
\( \mathcal{I} \equiv n_L - n_R\) where 
$n_L$ and $n_R$ are the zero modes of the left handed and 
the right handed fermions respectively. This index can be computed
using the formula \cite{Eweinberg}\cite{GanLaz}
\be
\mathcal{I} = \frac{1}{2\pi i}(\ln \textrm{det}M)\vert^{2\pi}_{\phi=0}
\label{eq:index}
\ee
where  $M$ is the position 
dependent effective mass matrix for the fermions. 

Thus, using 
either of  the results \cite{JandRossi} or \cite{GanLaz} i.e.,\  eq. 
(\ref{eq:index})
we can see that there will be $13$   zero modes present in the lowest 
winding sector of the cosmic string. Thus the induced fermion number 
differs from that of the vacuum by half-integer as required.
According to well known reasoning 
\cite{JandR} to be recapitulated 
below, this  requires the assignment of either of the values 
$\pm1/2$ to the fermion number of this configuration.

\section{Assignment of fermion number}\label{sec:ferno}
We now recapitulate the reasoning behind the assignment of
fractional fermion number. We focus on the Majorana fermion
case, which is more nettlesome, while the treatment of the
Dirac case is standard \cite{JandR}\cite{Jrev}. In the prime example 
in $3+1$ dimensions 
of a single left-handed fermion species $\Psi_L$ coupled to an abelian
Higgs model according to 
\be
\mathcal{L}_\psi = i\overline{\Psi_L}\gamma^\mu D_\mu \Psi_L
-\frac{1}{2}( h \phi \overline{\Psi_L^C}\Psi_L + h.c.)
\label{fermion_case_two}
\ee
the following result has been obtained\cite{JandRossi}.
For a vortex oriented along 
the $z$-axis, and in the winding number sector $n$, 
the fermion zero-modes are of the form
\be
\psi_0({\bf x})= \begin{pmatrix} 1\\ 0 \end{pmatrix}
\left[U(r)e^{il\phi}+V^*(r)e^{i(n-1-l)\phi}\right]g_l(z+t)
\ee
In the presence of the vortex, $\tau^3$ (here representing
Lorentz transformations on spinors) acts as the
matrix which exchanges solutions of positive frequency
with those of negative frequency. It is therefore identified
as the ``particle conjugation'' operator.
In the above ansatz, the $\psi$ in the zero-frequency 
sector are charge self-conjugates, $\tau^3\psi=\psi$,
and have an associated left moving zero mode along the
vortex. The functions satisfying $\tau^3\psi=-\psi$
are not normalizable. The situation is reversed 
when the winding sense of the scalar field is reversed, ie, for
$\sigma_u\sim$$e^{-in\phi}$. In the winding number sector $n$, 
regular normalizable solutions\cite{JandRossi} exist for
for $0\leq l\leq n-1$. The lowest energy sector of the vortex is 
now \(2^n\)-fold degenerate, and each zero-energy mode needs to be 
interpreted as contributing a value $\pm 1/2$ to the total 
fermion number of the individual states\cite{JandR}. This conclusion
is difficult to circumvent if the particle spectrum is to reflect
the charge conjugation symmetry of the theory \cite{sudyaj}. 
The lowest possible value of the induced number in this sector
is $-n/2$. Any general state of the system is built from one
of these states by additional integer number of fermions.
All the states in the system therefore possess half-integral
values for the fermion number if $n$ is odd. 

One puzzle immediately arises, what is the meaning of
negative values for the fermion number operator for
\textit{Majorana} fermions? In the trivial vacuum, we can
identify the Majorana  basis as
\be
\psi\ =\ \frac{1}{2}(\Psi_L +  \Psi_L^C)
\label{majdef}
\ee
This leads to the Majorana condition which results in
identification of particles with anti-particles according to
\be
\sC \psi \dsC\  =\   \psi 
\label{majcond}
\ee
making negative values for the number meaningless.
Here $\sC$ is the charge conjugation operator.  We shall 
first verify that in the zero-mode 
sector we must indeed assign negative values to the number 
operator. It is sufficient to treat 
the case of a single zero-mode, which generalizes easily
to any larger number of zero-modes.
The number operator possesses the properties
 \be
[ N, \psi ]\ =\ -\psi\qquad{\rm and} 
\qquad [ N, \psi^\dagger ]\ =\ \psi^\dagger
\label{psiN}
\ee
\be
\sC N \dsC\ =\ N
\label{ccN}
\ee
Had it been the Dirac case, there should be a minus sign 
on the right hand side of eq. (\ref{ccN}). This is absent due 
to the Majorana condition. The fermion field operator for the
lowest winding sector is now expanded as
\be
\psi\ =\ c\psi_0 + \left\{ \sum_{{\bf \kappa},s} 
a_{{\bf \kappa},s}\chi_{{\bf \kappa},s}(x)\ \\
+\ \sum_{{\bf k},s} b_{{\bf k},s}u_{{\bf k},s}(x)\ 
+\ h.c. \right\}
\label{eq:expansion}
\ee
where the first summation is over all the possible bound states of 
non-zero frequency with real space-dependence of the form $\sim
e^{-{\bf \kappa\cdot x_\perp}}$ in the transverse space directions
${\bf x}_\perp$, and the second summation is over all 
unbound states, which are asymptotically plane waves. 
These summations are suggestive and their exact connection to
the Weyl basis mode functions \cite{FukSuz}  are not essential 
for the present purpose.
Note however that no "$h.c.$" is needed for the zero energy mode
which is self-conjugate.  Then the Majorana condition (\ref{majcond}) 
requires that we demand
\be
\sC\ c\ \dsC\ =\   c \qquad{\rm and} 
\qquad \sC\  c^\dagger\  \dsC\ =\  c^\dagger
\label{ccc}
\ee
Unlike the Dirac case, the $c$ and $c^\dagger$ are not
exchanged under charge conjugation. 
The only non-trivial irreducible realization of this
algebra is to require the existence of a doubly degenerate
ground state with states $\state{-}$ and $\state{+}$ satisfying
\be
c\state{-}\ =\ \state{+}\qquad {\rm and}\qquad 
c^\dagger\state{+}\ =\ \state{-}
\label{cstates}
\ee
with the simplest choice of phases. Now we find
\begin{eqnarray}
\sC\  c\  \dsC \sC \state{-}\ &=\ \sC\state{+}\\ \vspace{3mm}
\Rightarrow \hspace{2mm}  c (\sC \state{-})\ &=\ (\sC\state{+})
\label{Ctransform}
\end{eqnarray}
This relation has the simplest non-trivial solution
\be
\sC\state{-}\ =\ \eta^-_{\ssC} \state{-}\qquad
{\rm and}\qquad \sC\state{+}\ =\ \eta^+_{\ssC} \state{+}
\label{Cproperty}
\ee
where, for the consistency of  (\ref{cstates}) and (\ref{Ctransform})
$\eta^-_{\ssC}$ and $\eta^+_{\ssC}$ must satisfy
\be
(\eta^-_{\ssC})^{-1}\eta^+_{\ssC}\ =\ 1
\ee
Finally we verify that we indeed get values $\pm 1/2$ for $N$.
The standard fermion number operator which in the Weyl basis is
\be
N_F = \frac{1}{2}[ \Psi_L^\dag \Psi_L - \Psi_L \Psi_L^\dag ]
\ee
acting on these two states gives,
\be
\frac{1}{2}(c\ c^\dagger\  -\  c^\dagger\  c)\ \state{\pm}\ =\ 
\pm\frac{1}{2}\state{\pm}
\ee
The number operator indeed lifts the degeneracy of the
two states. For $s$ number of zero modes, the ground
state becomes $2^s$-fold degenerate, and the fermion number
takes values in integer steps ranging from $-s/2$ to $+s/2$.
For $s$ odd the values are therefore half-integral.
Although uncanny, these conclusions accord with some known
facts. They can be understood as spontaneous symmetry breaking 
for fermions\cite{mcwilczek}. The negative values of the number thus
implied occur only in the zero-energy sector and do not 
continue indefinitely to $-\infty$. Instead of an unfathomable 
\textit{Dirac sea} we have a small \textit{Majorana pond} 
at the threshold.

\section{Energetics and dynamics of the thin, false string}
\label{sec:falsestring}

In the introductory section I mentioned the need for an instanton type 
description  for tunneling in the string induced decay of the false vacuum.  
It was only many years later, in collaboration with Montr{\'e}al and Seoul 
colleagues that this goal finally got taken up. While the insights were 
certainly new, there was a strong  impetus for completing the argument because 
of the tremendous growth and ease of use of computing powers. 

The work presented is a continuation of the earlier joint work 
\cite{Kumar:2010mv}\cite{Lee:2013ega}.  Here we  consider the case of cosmic 
strings in a spontaneously broken $U(1)$ gauge theory, a generalized Abelian 
Higgs model.    The potential for the complex scalar field  has a local minimum 
at a nonzero value  and the true minimum is at vanishing scalar field.  We 
assume the energy density splitting between the false vacuum and the true vacuum 
is very small. The spontaneously broken vacuum is the false vacuum.   
In the scenario that we have described, the true vacuum lies at the 
regions of vanishing scalar field, thus the interior of the cosmic string is in 
the true vacuum while the exterior is in the false vacuum.  

Related work of similar nature can be found in 
\cite{Kumar:2009pr}\cite{Aguirre:2009tp}. More recently this 
phenomenon has drawn attention in other field theoretic contexts 
\cite{Haberichter:2015xga}\cite{Dupuis:2015fza}, and in superstring theory, 
similar results are obtained regarding brane induced vacuum decay. See for 
instance \cite{Eto:2012ij}\cite{Kamada:2013rya}\cite{Kasai:2015maa} 
\cite{Kasai:2015dia}\cite{Kasai:2015exa}\cite{Lee:2015rwa}\cite{
Garbrecht:2015yza} \cite{Nakai:2016amt}.

\subsection{Set-up}

We consider the abelian Higgs model (spontaneously-broken scalar electrodynamics) with a 
modified scalar potential corresponding to our previous work \cite{Lee:2013ega} 
but now generalized to 
3+1 dimensions.  The Lagrangian density of the model has the form
\beq
{\cal L} = - \frac{1}{4} F_{\mu\nu}F^{\mu\nu} + (D_{\mu}\phi)^*(D^{\mu}\phi)-V(\phi^*\phi),
\label{lagran01}
\eeq
where
$F_{\mu\nu} = \partial_{\mu}A_{\nu} - \partial_{\nu}A_{\mu}$ and
$D_{\mu}\phi = (\partial_{\mu} - ie A_{\mu})\phi$.
The potential is a sixth-order polynomial in $\phi$
\cite{Kumar:2010mv, pjs}, written
\beq
V(\phi^*\phi) = \lambda(|\phi |^2-\eps v^2) (|\phi |^2-v^2)^2. \label{potential1}
\eeq
Note that the Lagrangian is no longer renormalizable in 3+1 dimensions, however 
the  understanding is that it is an effective theory obtained from a well 
defined renormalizable fundamental Lagrangian. The fields $\phi$ and $A_\mu$, 
the vacuum expectation value $v$ have mass dimension 1,  the charge $e$ is 
dimensionless and $\lambda$ has mass dimension 2 since it is the coupling 
constant of the sixth order scalar potential. The potential energy density of 
the false vacuum $|\phi |=v$ vanishes, while that of the true vacuum has 
$V(0)=-\lambda v^6\eps$.   We rescale analogous to \cite{Lee:2013ega} 
\beq
\phi\rightarrow v\phi\quad A_\mu\rightarrow vA_\mu \quad e\rightarrow\lambda^{1/2}ve\quad x\rightarrow x/(v^2\lambda^{1/2})
\eeq
so that all fields, constants and the spacetime coordinates become dimensionless, then the Lagrangian density is still given by Eqn. (\ref{lagran01}) where now the potential is 
\beq
V(\phi^*\phi) = (|\phi |^2-\eps) (|\phi |^2-1)^2. \label{potential}
\eeq
and there is an overall factor of $1/(\lambda v^2)$ in the action.

Initially, the cosmic string will be independent of $z$ the coordinate along its length and will correspond to a tube of radius $R$ with a trapped magnetic flux in the true vacuum inside, separated by a thin wall from the false vacuum outside.  $R$ will vary in Euclidean time $\tau$ and in $z$ to yield an instanton solution.  Thus we promote $R$ to a field $R\rightarrow R(z,\tau)$.  Hence we will look for axially-symmetric solutions for $\phi$ and $A_{\mu}$ in
cylindrical coordinates $(r$, $\theta$, $z$, $\tau)$. We use the following  ansatz
for a vortex of winding number $n$:
\beq
\phi(r, \theta, z, \tau) = f(r, R(z,\tau)) e^{in\theta}, \qquad  A_{i}(r, 
\theta ,z, \tau)=-\frac{n}{e} \frac{\varepsilon^{ij}{r}_j }{r^2}a(r,R( z,\tau)),
\label{ansatz}
\eeq
where $\varepsilon^{ij}$ is the
two-dimensional Levi-Civita symbol.  This ansatz is somewhat simplistic, it is clear that if the radius of the cosmic string swells out at some range of $z$, the magnetic flux will dilute and hence through the (Euclidean) Maxwell's equations some ``electric'' fields will be generated.  In 3 dimensional, source free, Euclidean electrodynamics, there is no distinct electric field, the Maxwell equations simply say that the 3 dimensional magnetic field is divergence  free and rotation free vector field that satisfies superconductor boundary conditions at the location of the wall.  It is clear that the correct form of the electromagnetic fields will not simply be  a diluted magnetic field that always points along the length of the cosmic string as with our ansatz, however the correction will not give a major contribution, and we will neglect it.  Indeed, the induced fields will always be smaller by a power of $1/c^2$ when the usual units are used.  

In the thin wall limit, the Euclidean action can be evaluated essentially 
analytically,  up to corrections which are smaller by at least one power of 
$1/R$.  The method of evaluation is identical to that in \cite{Lee:2013ega}, we 
shall not repeat the details, we find
\beq
S_E=\frac{1}{\lambda v^2}\int d^2x \frac{1}{2}M(R(z,\tau))(\dot R^2+R'^2) +E(R(z,\tau))-E(R_0)
\eeq
where
\bea
M(R)&=&\left[ \frac{2\pi n^2}{e^2R^2}+\pi R  \right] \\
E(R)&=&\frac{n^2\Phi^2}{2\pi R^2}+\pi R-\eps\pi R^2
\label{eq:MandE}
\eea
and $R_0$ is the classically stable thin tube string radius. 

\section{Instantons and the bulge}
\label{sec:instantonandbulge}

\subsection{Tunnelling instanton}
\label{sec:instanton}
We look for an instanton solution that is $O(2)$ symmetric, the appropriate ansatz is
\beq
R(z,\tau)=R(\sqrt{z^2+\tau^2})=R(\rho)
\eeq
with the imposed boundary condition that $R(\infty)=R_0$. 

Such a solution will describe the transition from a string of radius $R_0$ at $\tau=-\infty$, to a point in $\tau=\rho_0$ say at $z=0$ when a soliton anti-soliton pair is started to be created.  The configuration then develops a bulge which forms when the pair separates to a radius which has to be again $\rho_0$ because of $O(2)$ invariance and which is the bounce point of the instanton  along the $z$ axis at $\tau=0$.  Finally the  subsequent Euclidean time evolution continues in a manner which is just the  (Euclidean) time reversal of evolution leading up to the bounce point configuration  until a simple cosmic string of radius $R_0$ is re-established for $\tau\ge\rho_0$ and all $z$, {\it i.e.} $\rho\ge\rho_0$.  The action functional is given by
\beq
S_E=\frac{2\pi}{\lambda v^2}\int d\rho\,\, \rho \left[ \frac{1}{2}M(R(\rho))\left(\frac{\partial R(\rho)}{\partial\rho}\right)^2 +E(R(\rho))-E(R_0)\right] .
\eeq
The instanton equation of motion is
\beq
\frac{d}{d\rho}\left(\rho M(R)\frac{dR}{d\rho}\right) -\frac{1}{2}\rho M'(R)\left(\frac{dR}{d\rho}\right)^2-\rho E'(R)=0
\eeq
with the boundary condition that $R(\infty)=R_0$, and we look for a solution that has $R\approx R_1$ near $\rho =0$.  The solution necessarily ``bounces'' at $\tau=0$ since $\partial R(\rho)/\partial\tau|_{\tau=0}=R'(\rho) (\tau/\rho)|_{\tau=0}=0$. (The potential singularity at $\rho=0$ is not there since a smooth configuration requires $R'(\rho)|_{\rho =0}=0$.)  The equation of motion is better cast as an essentially conservative dynamical system with a ``time'' dependent mass and the potential given by the inversion of the energy function 
of Eq. (\ref{eq:MandE}), 
but in the presence of a ``time'' dependent friction where $\rho$ plays the role of time:
\beq
\frac{d}{d\rho}\left(M(R)\frac{dR}{d\rho}\right) - \frac{1}{2}M'(R)\left(\frac{dR}{d\rho}\right)^2- E'(R)=-\frac{1}{\rho}\left(M(R)\frac{dR}{d\rho}\right).\label{eqofm}
\eeq
As the equation is ``time'' dependent, there is no analytic trick to evaluating 
the bounce configuration and the corresponding action.  However, we can be 
resonably sure of the existence of a solution which starts with a given 
$R\approx R_1$ at $\rho=0$ and achieves $R=R_0$ for $\rho>\rho_0$, by showing 
the existence of an initial condition that gives an overshoot and another 
initial condition that gives an undershoot, in the same manner of proof as in 
\cite{Coleman}. 
 \begin{figure}[ht]
\centerline{\includegraphics[width=0.9\linewidth]{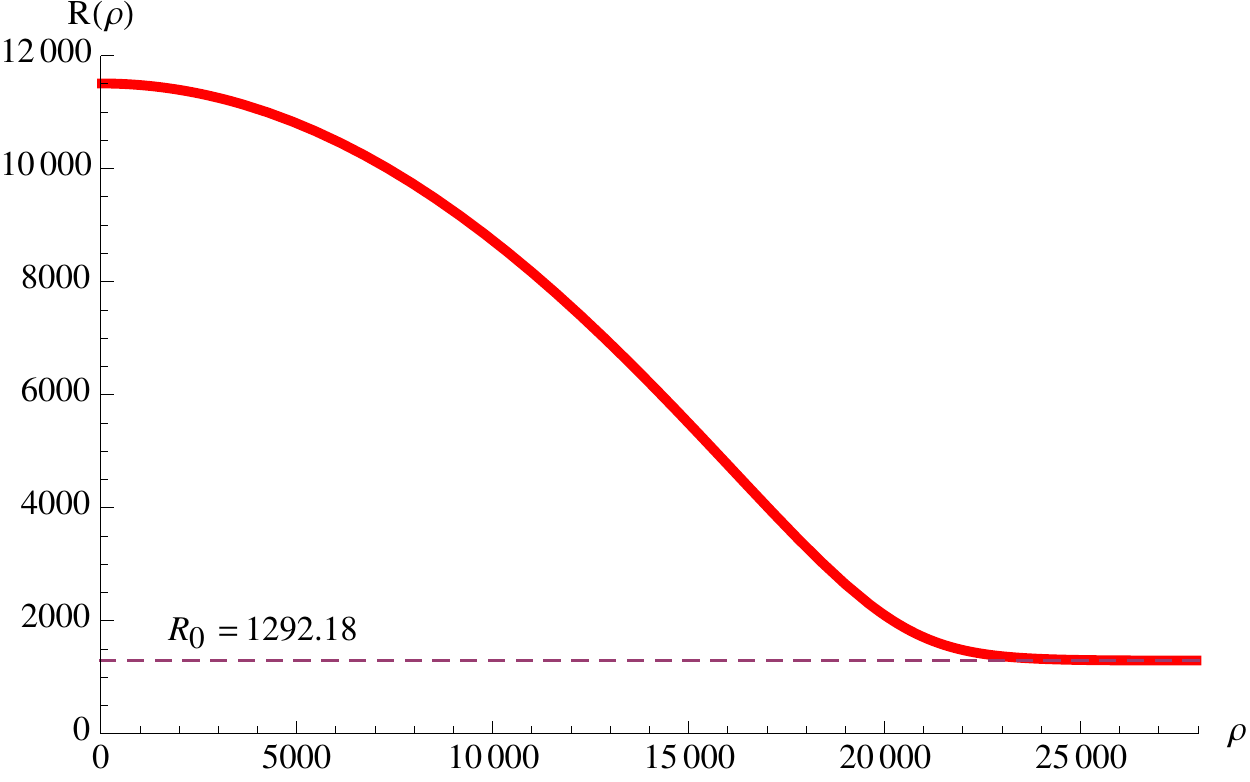}} 
\caption{ (color online)  The collective variable $R$ signifying the radius of 
the string in the thin wall approximation, as a function of 2-dimensional 
radius $\rho$ in the Euclidean $\tau-z$ plane.}
\label{radius}
\end{figure}
Actually, numerically integrating to $\rho\approx 80,000$ the function falls 
back to the minimum of the inverted energy functional Eq.
\ref{eq:MandE}.  On the other hand, we increase the starting point by 
$.0001$, the numerical solution overshoots the maximum at $R=R_0$.  Thus we have 
numerically implemented the overshoot/undershoot criterion of \cite{Coleman}.

\begin{figure}
\begin{center}
\includegraphics[width=10.5cm]{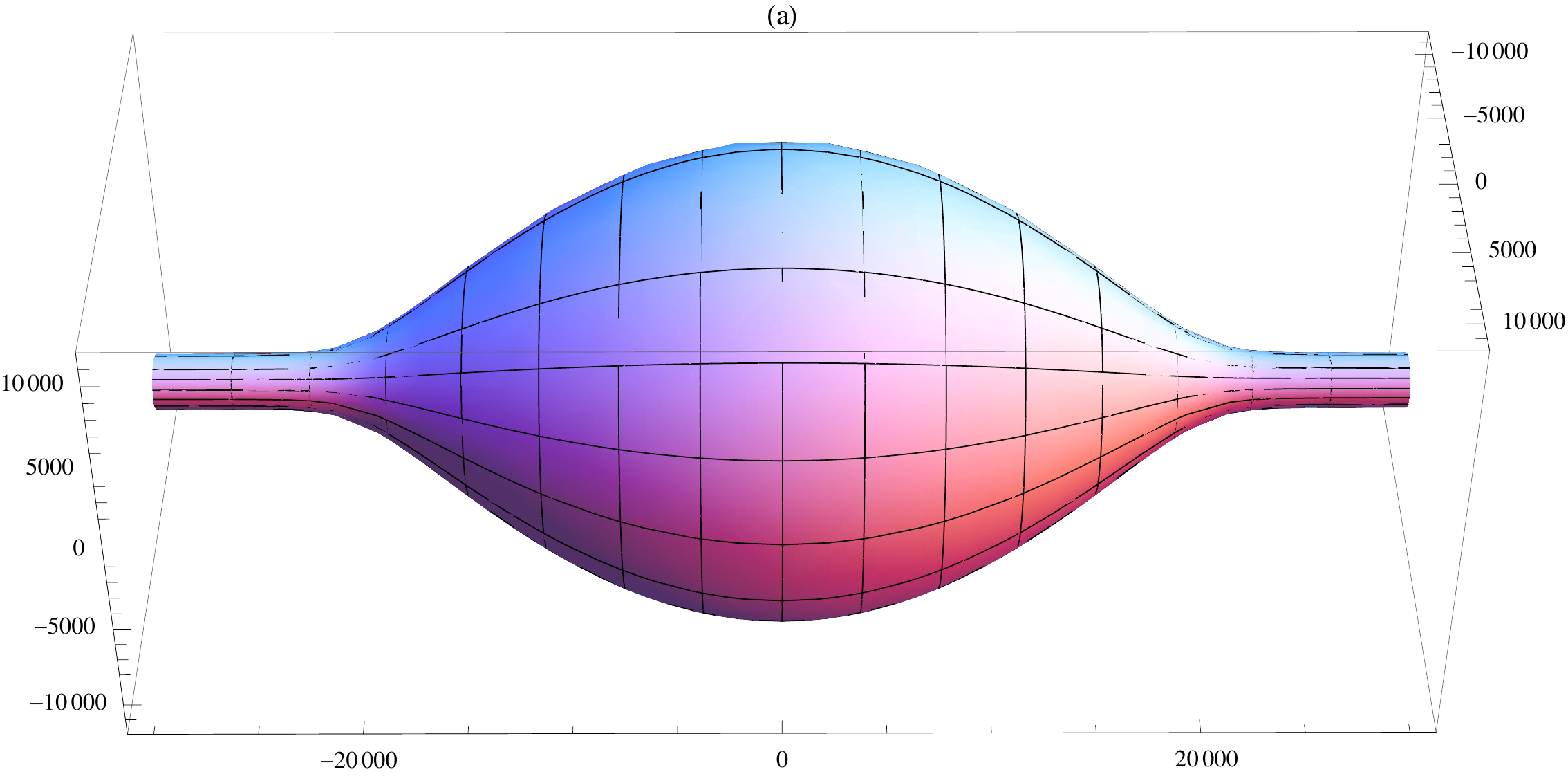}
\qquad
\includegraphics[width=12cm]{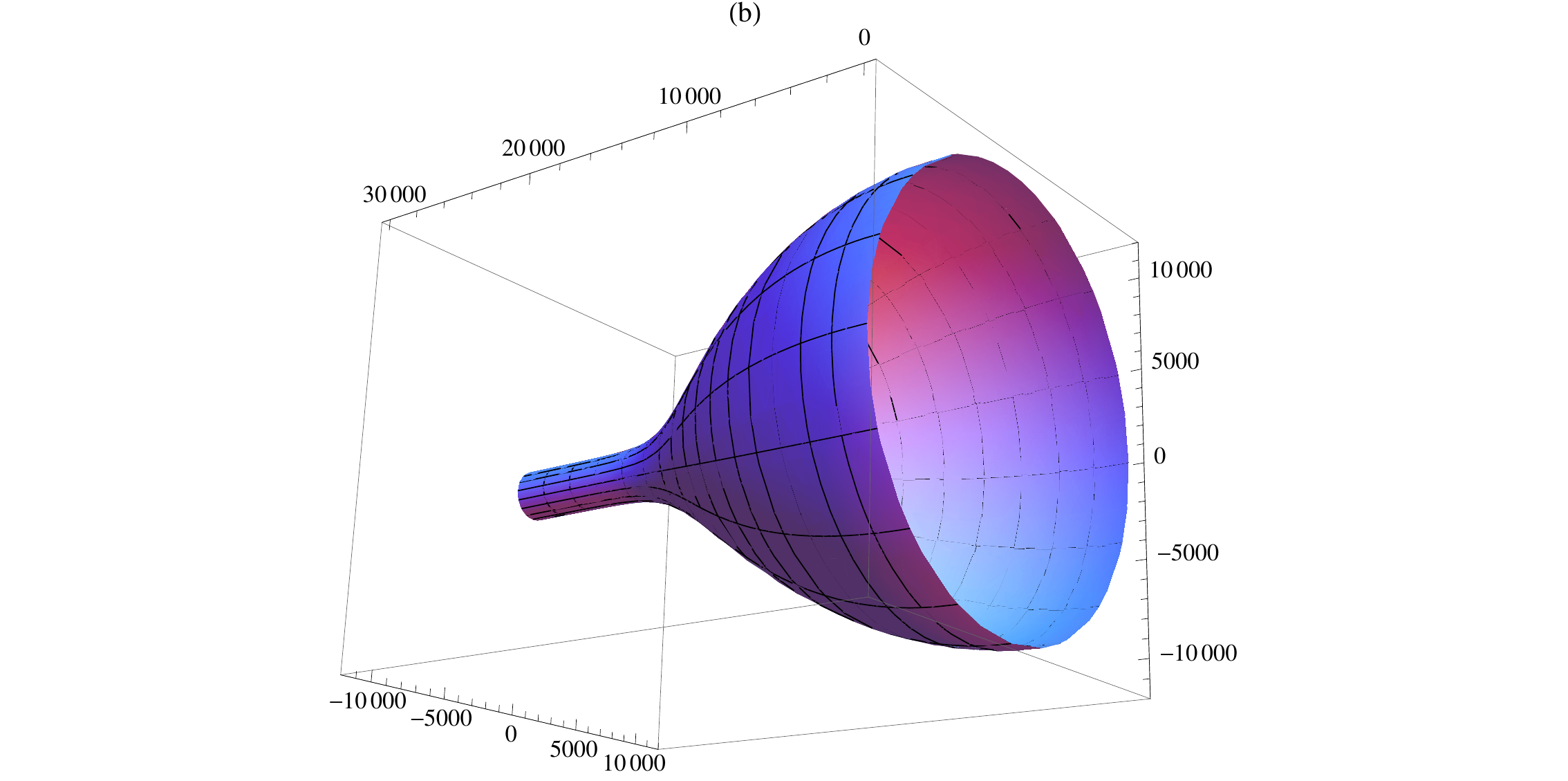}
\end{center}
\caption{(Color online) (a) Cosmic string  profile at  the bounce point.  (b) Cut away of the cosmic string profile at bounce point. } 
\label{bp}
\end{figure}

The cosmic string emerges with a bulge described by the function numerically evaluated 
and represented in Fig. \ref{radius} 
which corresponds to $R(z, \tau=0)$. A cross section of the bounce is 
visualised as the symmetrised figure obtained by reflecting the graph of Fig. 
\ref{radius} in the $R$ axis, with $-\infty < \rho < \infty$. 
A 3-dimensional depiction of the bounce point is given in Fig. \ref{bp}.  

This radius function has argument $\rho=\sqrt{z^2+\tau^2}$.  Due to the Lorentz invariance of the original action, the subsequent Minkowski time evolution is given by $R(\rho)\to R(\sqrt{z^2-t^2})$, which is of course only valid for $z^2-t^2\ge 0$. Fixed $\rho^2=z^2-t^2$ describes a space-like hyperbola that asymptotes to the light cone.  The value of the function $R(\rho)$ therefore remains constant along this hyperbola.  This means that the point at which the string has attained the large radius moves away from $z\approx 0$ to $z\to\infty$ at essentially the speed of light.  The other side of course moves towards $z\to -\infty$.  Thus the soliton anti-soliton pair separates quickly moving at essentially the speed of light, leaving behind a fat cosmic string, which is subsequently, classically unstable to expand and fill all space.  

\subsection{Tunnelling amplitude}
In the 3+1 dimensional example we have presented the bounce action needs 
to be found numerically. It 
is reasonable to expect that as $\eps\to 0$ the tunnelling barrier will get 
progressively bigger and at some point the tunnelling amplitude will vanish. On 
the other hand, there should exist a limiting value, call it $\eps_c$, where  
the tunnelling barrier disappears at the so-called dissociation point 
\cite{Yajnik:1986tg}\cite{Kumar:2008jb}\cite{Kumar:2009pr}, such that as 
 $\eps\to\eps_c$,  the action of the instanton will vanish, analogous to what 
was found in \cite{Lee:2013ega}. This is not possible to demonstrate in this 
case. However, for the bounce found above, we compare its implications to the 
other situations. Given the bounce action, the decay rate per unit 
length of the cosmic 
string will be of the form
\beq
\Gamma= A^{\rm c.s.} \left(\frac{S_0(\eps)}{2\pi}\right)e^{-S_0(\eps)}.
\eeq
where $A^{\rm c.s.}$ is the determinantal factor excluding the zero modes and $\left(\frac{S_0(\eps)}{2\pi}\right)$ is the correction obtained after taking into account the two zero modes of the bulge instanton.  These  correspond to invariance under Euclidean time translation  and   spatial translation along the cosmic string \cite{Coleman}. 
In general, there will be  a length $L$ of cosmic string per volume $L^3$.  For a second order phase transition to the metastable vacuum, $L$ is the correlation length at the temperature of the transition which satisfies $L^{-1}\approx \lambda v^2 T_c$ \cite{kz}.  For first order transitions, it is not clear what the density of cosmic strings will be.  We will keep $L$ as a parameter but we do expect that it is microscopic.  Then in a large volume $\Omega$, we will have a total length $N L$ of cosmic string, where $N=\Omega/L^3$.  Thus the decay rate for the volume $\Omega$ will be
\beq
\Gamma \times (NL)=\Gamma \left(\frac{\Omega}{L^2}\right)= A^{\rm c.s.} \left(\frac{S_0(\eps)}{2\pi}\right)e^{-S_0(\eps)}\frac{\Omega}{L^2}
\eeq
or the decay rate per unit volume will be
\beq
\frac{\Gamma}{L^2}=\frac{A^{\rm c.s.} \left(\frac{S_0(\eps)}{2\pi}\right)e^{-S_0(\eps)}}{L^2}.
\eeq
A comparable calculation with point-like defects \cite{Lee:2013ega} would give a 
decay rate per unit volume of the form
\beq
\frac{\Gamma^{\rm point\, like} }{L^3}=\frac{A^{\rm point\, like} \left(\frac{S_0^{\rm point\, like}(\eps)}{2\pi}\right)^{3/2}e^{-S_0^{\rm point\, like}(\eps)}}{L^3}
\eeq
and the corresponding decay rate from vacuum bubbles (without topological defects) \cite{Coleman} would be
\beq
\Gamma^{\rm vac.\, bubble} =A^{\rm vac.\, bubble} \left(\frac{S_0^{\rm vac.\, bubble} (\eps)}{2\pi}\right)^2e^{-S_0^{\rm vac.\, bubble} (\eps)}.
\eeq
Since the length scale $L$ is expected to be microscopic,  we would then find that the number of defects in a macroscopic volume ({\it i.e.} universe) could be incredibly large, suggesting that the decay rate from topological defects would dominate over the decay rate obtained from simple vacuum bubbles ˆ la Coleman \cite{Coleman}.   Of course the details do depend on the actual values of the Euclidean action and the determinantal factor that is obtained in each case.   

\section{Conclusion}

Metastable classical lumps, also referred to as embedded defects
can be found in several theories.
The conditions on the geometry of the vacuum manifold
that give rise to such defects were spelt out in \cite{presvil}.
We have studied the related question of fermion zero-energy 
modes on such objects. It is possible to construct examples
of cosmic strings in which the presence of zero-modes signals 
a fractional fermion number both for Dirac
and Majorana masses. 
It then follows that such a cosmic string cannot decay in
isolation because it belongs to a distinct superselected 
Quantum Mechanical sector. Thus a potentially metastable object 
can enjoy induced stability due to its bound state with fermions.

Although decay is not permitted in isolation, it certainly
becomes possible when more than one such objects come
together in appropriate numbers. In the early Universe such 
objects could have formed depending on the unifying group 
and its breaking pattern. Their disappearance would be
slow because it can only proceed through encounters between
objects with complementary fermion numbers adding up to an 
integer. Another mode of decay is permitted by change in
the ground state in the course of a phase transition.
When additional Higgs fields acquire a vacuum expectation 
value, in turn altering the boundary conditions for the
Dirac or Majorana equation, the number of induced zero
modes may change from being odd to even thus imparting
the strings an integer fermion number. The decay can then proceed
at the rates calculated in \cite{presvil}. Such a possibility,
for the case of topologically stable string can be found for realistic
unification models in \cite{steandyaj} \cite{Davis:1997bu} 
\cite{Davis:1999ec} \cite{widyajetal}. 

There are many instances where the vacuum can be meta-stable.  The symmetry broken vacuum can  be  metastable.  Such solutions for the vacuum can be important for cosmology and for the case of supersymmetry breaking see \cite{abel} and the many references therein.  In string cosmology, the inflationary scenario that has been obtained in\cite{kklt},   also gives rise to a vacuum that is  meta-stable , and it must necessarily be long-lived to have cosmological relevance.  

In a condensed matter context symmetry breaking ground states are also of great importance.  For example, there are two types of superconductors \cite{am}. The cosmic string is called a vortex line solution in this context, and it is relevant to type II superconductors.   The vortex line contains an unbroken symmetry region that carries a net magnetic flux, surrounded by a region of broken symmetry.  If the temperature is raised, the true vacuum becomes the unbroken vacuum, and it is possible that the system exists in a superheated state where the false vacuum is meta-stable \cite{dolgert}.  This technique has actually been used to construct detectors for particle physics \cite{pd}.  Our analysis might even describe the decay of vortex lines in superfluid liquid 3Helium \cite{legg}.

The decay of all of these metastable states could be described through the tunnelling transition mediated by instantons in the manner that we have computed in this article.  For appropriate limiting values of the parameters, for example when $\eps\to\eps_c$, the suppression of tunnelling is absent, and the existence of vortex lines or cosmic strings could cause the decay of the meta-stable vacuum without bound.  Experimental observation of this situation would be interesting.

\begin{acknowledgement}
I thank the editors of this  volume for giving me the opportunity to 
contribute. This article is patched together from \cite{Sahu:2004zp} and 
\cite{Lee:2013zca}, and it is a pleasant duty to acknowledge the partnership of 
the coauthors of both the papers where the work was originally reported. I also 
acknowledge the support of the Department of Science and Technology, India, and 
the Minist\`{e}re des relations internationales du Qu\'{e}bec.
\end{acknowledgement}


\begin{thebibliography}{00}

\bibitem{Yajnik:1986tg}
  U.~A.~Yajnik,
  Phys.\ Rev.\ D {\bf 34} (1986) 1237.
  
\bibitem{Yajnik:1986wq}
  U.~A.~Yajnik and T.~Padmanabhan,
  Phys.\ Rev.\ D {\bf 35} (1987) 3100.

\bibitem{Yajnik:1990un}
  U.~A.~Yajnik,
  Phys.\ Lett.\ B {\bf 234} (1990) 271.
  
\bibitem{Sahu:2004zp}
  N.~Sahu and U.~A.~Yajnik,
  Phys.\ Lett.\ B {\bf 596} (2004) 1  [hep-th/0405140].

\bibitem{Lee:2013zca}
  B.~H.~Lee, W.~Lee, R.~MacKenzie, M.~B.~Paranjape, U.~A.~Yajnik and D.~h.~Yeom,
  Phys.\ Rev.\ D {\bf 88} (2013) no.10,  105008
  [arXiv:1310.3005 [hep-th]].

\bibitem{mcwilczek}
R.~MacKenzie and F.~Wilczek,
Phys.\ Rev.\ D {\bf 30}, 2194 (1984)

\bibitem{Lee:2013ega}
  B.~H.~Lee, W.~Lee, R.~MacKenzie, M.~B.~Paranjape, U.~A.~Yajnik and D.~h.~Yeom,
  Phys.\ Rev.\ D {\bf 88} (2013) 085031

\bibitem{3} H.~B.~Nielsen and P.~Olesen, Nucl.\ Phys.\ {\bf B61}, 45 (1973).
\bibitem{4} A. Abrikozov, JETP(Sov.Phys.), {\bf 5}, 1173 (1957).
  
\bibitem{JandR} R. Jackiw and C. Rebbi, Phys Rev. {\bf D13}, 3398 (1976)

\bibitem{Jrev} R. Jackiw, Rev. Mod. Phys. 49, 681(1977) 


\bibitem{devega} H. de Vega, Phys. Rev. {\bf D18}, 2932 (1978)

\bibitem{stern} A. Stern; Phys. Rev. Lett.{\bf 52}, 2118(1983)

\bibitem{steandyaj} A. Stern and U. A. Yajnik, Nucl. Phys. {\bf B267}, 158  (1986)

\bibitem{Davis:1997bu}
S.~C.~Davis, A.~C.~Davis and W.~B.~Perkins,
Phys.\ Lett.\ B {\bf 408}, 81 (1997) 

\bibitem{widyajetal} U.~A.~Yajnik, H.~Widyan, D.~Choudhari, S.~Mahajan 
and A.~Mukherjee, Phys.\ Rev.\ D {\bf 59}, 103508 (1999)

\bibitem{Davis:1999ec}
S.~C.~Davis, W.~B.~Perkins and A.~C.~Davis,
Phys.\ Rev.\ D {\bf 62}, 043503 (2000)

\bibitem{Davis:1999wq}
A.~C.~Davis, S.~C.~Davis and W.~B.~Perkins, \textsl{Proceedings of
COSMO99}, 
arXiv:hep-ph/0005091

\bibitem{DavKibetal}
A. C. Davis, T. W. B. Kibble, M. Pickles
and D. A. Steer, Phys. Rev. {\bf D62} 083516 (2000)


\bibitem{www} G. C. Wick, A. S. Wightman and E. P. Wigner, Phys. Rev. 
{\bf 88}, 101  (1952)

\bibitem{Sweinberg} S. Weinberg, {\it The Quantum Theory of fields}, Vol. I, 
sec. 3.3,  Cambridge University Press, 1996

\bibitem{sudyaj} E. C. G. Sudarshan and U. A. Yajnik, Phys. Rev. {\bf D33}, 
1830 (1986)

\bibitem{presvil} J. Preskill and A. Vilenkin, Phys. Rev. {\bf D47}, 
2324 (1993)

\bibitem{Eweinberg} E. J. Weinberg, Phys. Rev. {\bf D24}, 2669 (1981)

\bibitem{GanLaz} N. Ganoulis and G. Lazarides, Phys. Rev.
{\bf D38} 547 (1988)

\bibitem{Slansky:yr}
R.~Slansky,
Phys.\ Rept.\  {\bf 79}, 1 (1981).

\bibitem{JandRossi} R. Jackiw and P. Rossi, Nucl. Phys.{\bf B190}, 681 (1981)

\bibitem{DHN} R. F. Dashen, B. Hasslacher and A. Neveu 
Phys.\ Rev.\ D {\bf 10}, 4138 (1974) 

\bibitem{nohl} C. R. Nohl, Phys.\ Rev.\ D  {\bf  12}, 1840 (1975) 

\bibitem{FukSuz} See for instance M.~Fukugita and T.~Yanagida, in 
"\textsl{Physics and Astrophysics of Neutrinos}", 
M.~Fukugita and A.~Suzuki, (eds.), Springer-Verlag, (1994), pp. 1-248

\bibitem{Jackiw:wc}
R.~Jackiw and J.~R.~Schrieffer,
Nucl.\ Phys.\ B {\bf 190} (1981) 253.



\bibitem{Kumar:2010mv}
  B.~Kumar, M.~B.~Paranjape and U.~A.~Yajnik,
  Phys.\ Rev.\ D {\bf 82} (2010) 025022


\bibitem{Kumar:2009pr}
  B.~Kumar and U.~A.~Yajnik,
  Nucl.\ Phys.\ B {\bf 831} (2010) 162

\bibitem{Aguirre:2009tp}
  A.~Aguirre, M.~C.~Johnson and M.~Larfors,
  Phys.\ Rev.\ D {\bf 81} (2010) 043527

\bibitem{Haberichter:2015xga}
  M.~Haberichter, R.~MacKenzie, M.~B.~Paranjape and Y.~Ung,
  J.\ Math.\ Phys.\  {\bf 57} (2016) no.4,  042303


\bibitem{Dupuis:2015fza}
  \'{E}.~Dupuis, Y.~Gobeil, R.~MacKenzie, L.~Marleau, M.~B.~Paranjape and 
Y.~Ung,
  Phys.\ Rev.\ D {\bf 92} (2015) no.2,  025031

\bibitem{Eto:2012ij}
  M.~Eto, Y.~Hamada, K.~Kamada, T.~Kobayashi, K.~Ohashi and Y.~Ookouchi,
  JHEP {\bf 1303} (2013) 159
 
\bibitem{Kamada:2013rya}
  K.~Kamada, T.~Kobayashi, K.~Ohashi and Y.~Ookouchi,
  JHEP {\bf 1305} (2013) 091

\bibitem{Kasai:2015maa} 
  A.~Kasai, Y.~Nakai and Y.~Ookouchi,
  JHEP {\bf 1606}, 029 (2016)

\bibitem{Kasai:2015dia} 
  A.~Kasai and Y.~Ookouchi,
  JHEP {\bf 1506}, 098 (2015)

\bibitem{Kasai:2015exa} 
  A.~Kasai and Y.~Ookouchi,
  Phys.\ Rev.\ D {\bf 91}, no. 12, 126002 (2015)

\bibitem{Lee:2015rwa}
  B.~H.~Lee, W.~Lee and D.~h.~Yeom,
  Phys.\ Rev.\ D {\bf 92} (2015) no.2,  024027

\bibitem{Garbrecht:2015yza}
  B.~Garbrecht and P.~Millington,
  Phys.\ Rev.\ D {\bf 92} (2015) 125022
  
\bibitem{Nakai:2016amt} 
  Y.~Nakai and Y.~Ookouchi,
  arXiv:1608.01232 [hep-th].

\bibitem{pjs} P.~J.~Steinhardt, Phys.\ Rev.\ D {\bf 24}, 842 (1981).
\bibitem{Coleman} S. R. Coleman, Phys. Rev. D15, 2929 (1977); C. A. Callan and S. Coleman, Phys. Rev D16, 1762 (1977).

\bibitem{Kumar:2008jb}
  B.~Kumar and U.~A.~Yajnik,
  Phys.\ Rev.\ D {\bf 79} (2009) 065001

\bibitem{kz} T.~W.~B.~Kibble, J.~Phys.~A {\bf 9}, 1387 (1976); T.~W.~B.~Kibble, Phys. Rep. 67, 183 (1980); W.H. Zurek, Nature (London) 317, 505 (1985); W.H. Zurek, Acta Phys. Pol. B 24, 1301 (1993).
\bibitem{abel}S.~A.~Abel, C.~-S.~Chu, J.~Jaeckel and V.~V.~Khoze,  JHEP {\bf 0701}, 089 (2007) [hep-th/0610334]; Willy Fischler, Vadim Kaplunovsky, Lorenzo Mannelli, Marcus Torres and Chethan Krishnan, JHEP03(2007)107.
\bibitem{kklt} S. Kachru, R. Kallosh, A. D. Linde, and S. P. Trivedi, Phys. Rev. D68, 046005 (2003), hep-th/0301240.
\bibitem{am} N. W. Ashcroft and N. D. Mermin, Solid state physics,  (Harcourt College Publishers, 1976).
\bibitem{dolgert} A. J. Dolgert, S. J. Di Bartolo, and A. T. Dorsey, Phys. Rev. B 53, 5650, (1996).
\bibitem{pd} Superconductive Particle Detectors, edited by A. Barone (World Scientific, Singapore, 1987); K. Pretzl, J. Low Temp. Phys. 93, 439 (1993).
\bibitem{legg}A. J. Leggett, ÒA Theoretical Description of the New Phases of Liquid 3HeÓ, Rev. Mod. Phys. 47, 331, 1975.

\end{thebibliography}
\end{document}